# CloudPSS: A High-Performance Power System Simulator Based on Cloud Computing

Yankan Song[1,2], Ying Chen[1,2], Zhitong Yu[1,2], Shaowei Huang[1,2], Chen Shen[1,2]
[1]Dept. of Electrical Engineering, Tsinghua University
[2]Research Center of Cloud Simulation and Intelligent Decision-making, EIRI, Tsinghua University
Beijing, China
songyk@mail.tsinghua.edu.cn

*Abstract*—With the increasing computations in power system simulations, high-performance and cost-effective power system simulator is highly required. In this paper, a cloud-computing based power system simulator, namely CloudPSS, is designed. Based on the proposed open service integration framework, a self-developed electromagnetic transients (EMT) simulator with an automatic code generator is provided to accelerate EMT simulations using heterogeneous devices in the cloud, such as CPU and GPU. Test results show that CloudPSS can achieve significant speedups for both large-scale simulation tasks and multi-scenario tasks. Moreover, benefit from the time-sharing of cloud computing resources, the computational cost can be greatly reduced.

## I. INTRODUCTION

The rapid increase of power electronic devices has led to various complicated transient phenomena in modern power system. To improve system reliability and security, detailed simulations are required, whose accuracy and efficiency are related to modeling complexity, frequency span, system size and the number of contingencies considered. Especially, for analyzing fast transients in AC/DC hybrid power systems with renewable generations, the electromagnetic transient (EMT) simulation is of irreplaceable importance [1], [2].

However, EMT simulation for system-level applications may involve intensive computations and be time-consuming, since tiny time steps are adopted in the numeric integration of fast transient dynamics. For a large-scale AC/DC hybrid system, the computations are much larger in an EMT simulation than a transient stability program. Moreover, to perform multi-scenario studies, such as EMT simulation-based single outage (N-1) contingency analysis, the total time-costs are almost unbearable. As a result, to accelerate such applications, high-performance computing platforms are highly required.

Along with the development of computing technology in the past decades, researchers have proposed a variety of parallel methods on different computing devices to accelerate the EMT simulations [3]–[6]. A few of simulation tools have already supported preliminary parallel techniques using multi-core CPUs or PC clusters, such as the *IncrediBuild-XGE* in PSCAD [7]. However, most of the methods, such as those implemented on heterogeneous devices (e.g. GPU [5], [7]and FPGA[8]), are still difficult to deploy at the user side, as they always rely on special devices, drivers, compilers, and cumbersome configurations. Moreover, accelerations achieved by the parallel simulations generally depend on the computing resource configuration, such as the number of computing cores. As expansions of computing resources are expensive, normal users cannot afford to scale up the simulation capability arbitrarily.

During these years, the availability of high-capacity networks and high-performance computers have led to explosive growth in cloud computing [9]. Traditional offline software began to upgrade to the cloud, using cloud computing resources to provide services flexibly, i.e. Software-as-a-Service (SaaS). Based on the virtualization technique [10], the SaaS providers allow the users to concentrate on their core business without worrying about the computing resource bottlenecks. In addition, by time-sharing of computing cores, users can utilize various types of computing devices to obtain higher efficiency, while the computational cost of a single user is greatly reduced. Therefore, it is an easier-to-use and cost-effective way to build a cloud computing-based simulation platform.

In this paper, a self-developed high-performance EMT simulator on the cloud, or CloudPSS [11], is introduced. It is deployed on a heterogeneous cloud computing (HCC) platform [12], which adopts both CPU and GPU as computing devices. First, a cloud computing simulation platform based on virtualization technique is introduced. The simulation program for each task is automatically generated and packaged as an independent virtual simulation engine (VSE). Then, an automatic simulation code generator is designed to generate VSE according to the computation graph and computing device types. In addition, fine-grained parallel strategies proposed in [7] are adopted to optimize computing resources utilization in the VSE for both large-scale system tasks and multi-scenario studies.

The CloudPSS platform is designed to provide high-performance EMT simulation functions to end-users. It

This work was supported by the National Natural Science Foundation of China (Grant No. 51877115).

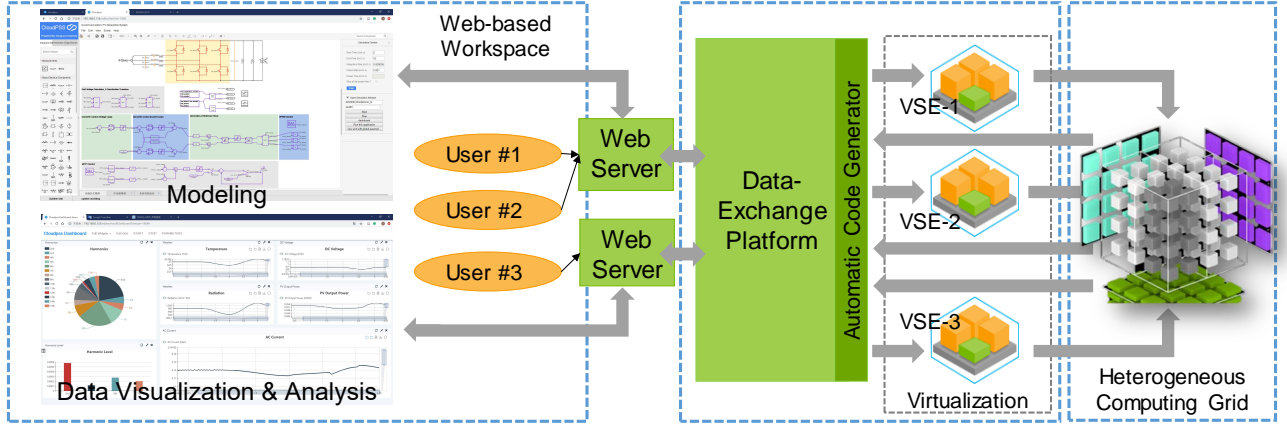

Figure 1 Open Cloud Service Integration Framework

combines the advantages of cloud computing and heterogeneous parallel computing. The features of this cloud-based simulator can be summarized as followings.

- **High efficiency**. Since the HCC platform comprises large amounts of CPU and GPU devices, parallel technologies [3]–[6], [13] can be deployed to speed up the simulation effectively.
- **Flexible-to-use**. Users only need to use the web browser without having to configuring complicate drivers, compilers and hardware environments.
- **Cost-effective**. Instead of maintaining local apparatus, the users shift to cloud computing resources to conduct simulation tasks. This cloud-based simulation environment tends to be inexpensive and highly scalable.

The rest of the paper are organized as follows. Section II briefly describes the framework of the CloudPSS platform. Section III introduces the automatic code generator in CloudPSS. Performance tests and results for large-scale cases as well as multi-scenario studies are provided in Section IV. The conclusions are given in Section V.

## II. HETEROGENEOUS CLOUD COMPUTING FRAMEWORK OF CLOUDPSS

As shown in Fig. 1, the CloudPSS platform is implemented in an open cloud service integration framework. It comprised of three subsystems, such as the web-based workspace, data-exchange platform, and parallel simulation engine. All the three parts are deployed in a third-party cloud service provider, such as AWS, Alibaba Cloud, Azure, and etc.

### A. Web-based Workspace

As shown in Fig. 1, the web-based workspace is a visualized modeling and data analysis tool developed mainly using HTML5 and web-socket technologies. The component models and data analyzing functions are provided as plugins. Users can construct electrical and control system models by dragging, dropping and connecting components. Models and simulation tasks are stored in JSON format, which can be converted from Simulink or PSCAD projects flexibly.

After the simulation model is completed, users can set up simulation tasks, configure waveform channels, and analyze results in the data visualization workspace. As all the above services are provided by a web server on cloud, the modeling workspace can be upgraded silently and smoothly. When new models and applications are published, they will become new plugins available for all users. That is, the user can always work on the latest version of CloudPSS tools without any local software installations.

### B. Data-exchange platform

The data-exchange platform is the middleware of CloudPSS. It mainly responsible for three functions, i.e., to generate virtual simulation engines (VSE) for each simulation task, to launch the VSE onto the heterogeneous computing grid and to store the results received from the computing grid.

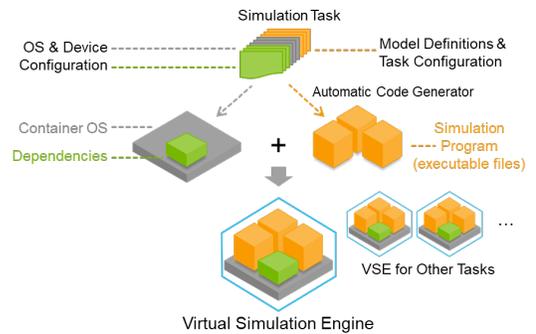

Figure 2 Automatic Generation of Virtual Simulation Engines

The data-exchange platform adopts virtualization technique to manage simulation tasks. Each simulation task is packaged as a Docker container [10], i.e., the virtual simulation engine (VSE). The VSE is regarded as a virtual machine containing the model data, EMT simulation program and its runtime dependencies.

When a simulation task is received, the data-exchange platform first selects a basic container as the VSE template. The basic container is comprised of a corresponding operating system, hardware driver and other dependencies, which are specified in the simulation task configurations. Then, the model definitions and task configurations are parsed by an automatic code generator. It automatically assembles and the EMT simulation program and compiles it to binary files, whose details are introduced in Section III. Finally, the simulation program with its basic container is packaged as a VSE and sent to the heterogeneous computing grid to carry out the simulation.

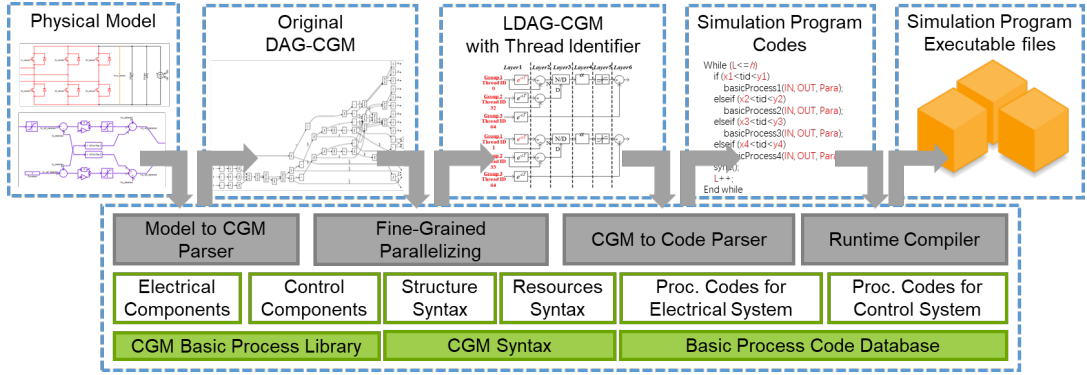

Figure 3 Automatic Code Generator based on Computation Graph

## C. Heterogeneous Computing Grid

The CloudPSS platform adopts heterogeneous computing devices, including x86-based CPU and CUDA-enabled GPU [14]. As a result, almost all the third-party cloud service providers are available to perform simulations. Especially, if the heterogeneous computing grid is deployed in a supercomputing platform with CUDA-enabled GPUs, the simulations will achieve significant speedups for large-scale system tasks and multi-scenario studies.

It should be noted that the proposed framework provides a comprehensive solution for increasing the security and scalability of the cloud-based simulator. First, VSEs are assembled for simulation tasks and independent from each other. Their failures will not interrupt the overall services of the cloud-based simulator. Then, as user-defined models and power grids are encapsulated into the VSE, which is compiled to binary codes, valuable designs and studies are only accessible and readable by their owners. Moreover, the VSE can be distributed to suitable computing resources flexibly and dynamically. Then, according to the efficiency preference of the user, multiple VSEs can be created and assigned to computing servers having different parallel computing equipment. All these computing resources are utilized by this certain user only temporarily. After all of the related VSEs finish their tasks, computing resources can be reused by other VSEs belong to other users. Thus, the scalability of the high-performance simulator can be enhanced significantly and inexpensively.

## III. AUTOMATIC CODE GENERATOR FOR HETEROGENEOUS COMPUTING DEVICES

In order to be compatible with various heterogeneous computing devices and to pursue higher computational efficiency, the simulation program needs to be reorganized and compiled at runtime according to the simulation model, device type, and the available resources. Therefore, the core computing program in the VSE is formed by an automatic code generator.

As shown in Fig. 3, the overall workflow to generate the simulation program includes three stages. First, the original model definitions and simulation configurations are translated into a computation graph model (CGM) [15]. Second, the computations inside the CGM are re-arranged using fine-grained parallel strategies [6], [16], while the computing resources are finely mapped into the CGM. Finally, the codes to proceed the CGM is automatically assembled and compiled to executable files on heterogeneous devices.

### A. Generation of Computation Graph Model

To organize the computations of the EMT simulation task more conveniently, the CloudPSS platform adopts a generalized computation graph model (CGM) which is proposed in [15]. The CGM method has been utilized to organize control system simulation and power flow computations in [6], [17], [18]. It has been proved to be highly efficient and flexible on heterogeneous platforms.

Typically, a CGM is defined as a directed graph $G = (V, E)$ of computations. The vertices inside a CGM represent basic processes, while the edges represent the data flow and interdependencies between vertices. A basic process can be regarded as one or several operations which perform sequentially, as shown in Fig. 4.

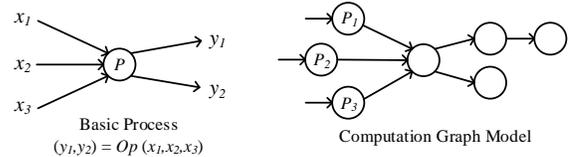

Figure 4 Basic Process (Vertice) and the Computation Graph Model

The method of expressing a control system using CGM has been provided in [7]. That is, each control component in the control system can be represented as a basic process, so the control system simulation model can be directly translated into a CGM. Similarly, in the EMTP-type simulation algorithm [2], the computations for the electrical system can also be represented by the following four kinds of basic processes:

- Updating Norton equivalent current and conductance for each electrical component;
- Updating nodal injected currents for each electrical component;
- Factorizing the system conductance matrix;
- Performing sparse forward and backward substitution to calculate nodal voltages.

Then, according to the type and quantity of electrical components in the original model definitions, the four basic processes can be organized as a CGM according to the EMTP-type algorithm.

In summary, when a simulation task is received by the automatic code generator, the model definitions are analyzed by a task parser. Using predefined process library and description syntax inside the automatic code generator, each EMT simulation task can be converted to a CGM. The automatic code generator further breaks all the algebraic loops inside the CGM, layer the vertices, and finally forms a layered directed acyclic graph-based CGM (LDAG-CGM).

### B. Fine-grained Parallel Computing for the CGM

If CUDA-enabled GPU is selected as computing device, the LDAG-CGM is then re-arranged using fine-grained parallel strategies to obtain higher efficiency and optimize the computing resources. Here, different parallel strategies are adopted for single large-scale simulation tasks and multi-scenario studies respectively.

- For a single large-scale simulation task, the computing resources are allocated to maximize the degree of parallelism for the LDAG-CGM. That is, by grouping the basic processes of each layer in the LDAG-CGM, all the computations of a layer are forced to execute in several single-instruction-multiple-threads (SIMT) [14] groups, which can be executed fully concurrently. The entire LDAG-CGM is then performed according to the layer order repeatedly.
- For a multi-scenario simulation task, the computing resources are allocated to maximize the core utilization rate using vectorization technique [19]. As different sub-scenarios in a multi-scenario task share the same system topology and simulation algorithm, so the LDAG-CGMs for all the sub-scenarios are isomorphic. Therefore, multi-scenario tasks can be accelerated by in a fully vectorized way. That is, the calculation of each basic process is extended to an SIMT operation to calculate all scenarios simultaneously.

In CUDA-enabled GPUs, the computing resources can be controlled by thread identifier [6], [19]. Therefore, after applying the fine-grained parallel acceleration strategy, the basic processes of each layer are reordered and grouped to form the final CGM description file. In addition, the thread identifier for each basic process is also stored in the CGM description file.

### C. Automatic Code Generation and Runtime Compiling

After the CGM is formed, it is analyzed by a CGM parser to generate the final simulation program for different devices. First, the CGM parser re-analyzes the CGM description file, allocate the memory space for each basic process. Then, it retrieves the codes for each basic process from the code database according to the type of computing device, and form the codes to calculate the entire CGM. Finally, the codes are compiled into executable files and assembled into the virtual simulation engine.

## IV. PERFORMANCE TESTS AND RESULTS

### A. Test Case and Environment

The CloudPSS test workspace is deployed on Alibaba Cloud [20]. Three type of processors are selected as computing devices, i.e., Intel E5-2682, NVIDIA K20x and P100. The VSEs are packaged in form of Docker containers.

A 33-node distribution network with photovoltaic subsystems [17] is selected as the basic test case, as in Fig. 5.

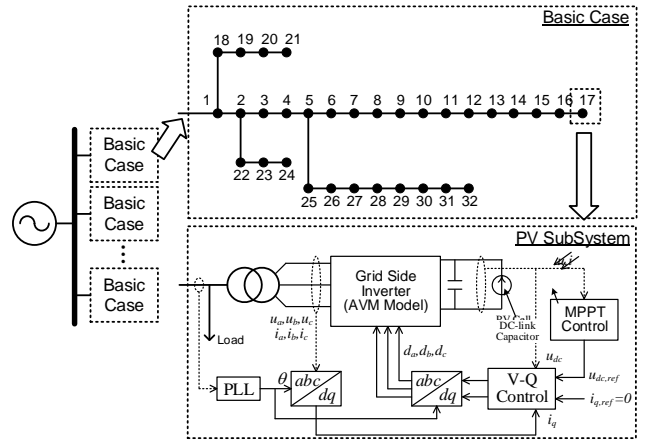

Figure 5 Topology of the Test Case.

### B. Performance of Large-scale Case

Large-scale cases are generated by connecting several basic cases at their root node (Node No. 1). The average time-costs of each integration step is measured and compared on different computing resources.

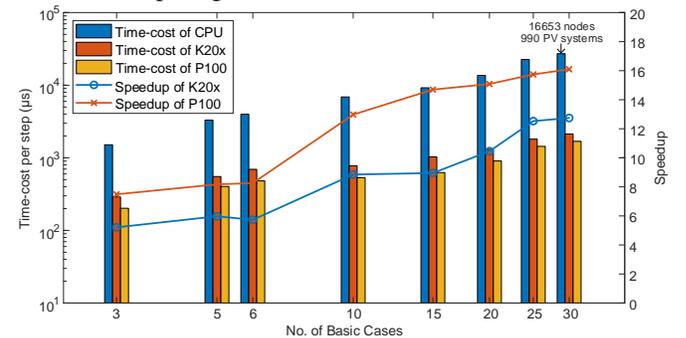

Figure 6 Time-costs and Speedup Performance for Large-scale Test Cases.

As can be seen from Fig. 6, the VSEs on both GPU platforms provide significant speedups compared to those on the CPU platform. In addition, the speedup ratio gradually increases as the scale of the system increases. For the largest system with 16,653 electrical nodes and 990 PV subsystems (77220 control components), the P100 platform offers 16x speedup than the CPU platform.

## C. Performance of Multi-Scenario Studies

The basic case with 3 PV subsystems is chosen for this test. More than 1000 scenarios are generated by changing the radiation and temperature of the PV panels. The average time-costs for all the scenarios per each integration step are measured and compared in Fig. 7.

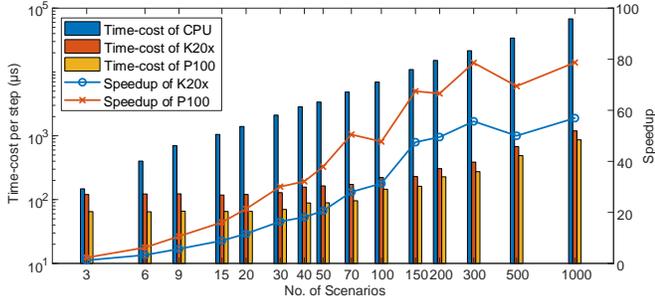

Figure 7 Time-costs and Speedup Performance for Multi-scenario Tasks.

Benefit from fully vectorized parallel strategies, the multi-scenario simulations on CloudPSS provides more significant speedups than the large-scale system simulations. It can be seen from Fig. 7 that before 40 scenarios, since the occupancy rate of the GPU has not reached 100%, the time-costs of VSEs on GPU platforms are almost unchanged. When launching 1000 scenarios, the VSE on P100 platform can provide up to 80x speedup compared to a fully serialized VSE on the CPU.

Considering that only a single device is used in each simulation test, in practical applications, users can rent more cloud computing resources to obtain higher computing efficiency. For example, to simulate the above 1000 scenarios, if 32 P100 computing devices are rented at the same time, fully parallelism can be achieved.

## D. Computational Cost Estimation

Using the CPU and GPU cloud computing resources provided by Alibaba Cloud [20], the average rental cost to finish 60s-physical-process simulations for 1000 scenarios are estimated according to the performance test results. As shown in Table 1, the GPU platform (P100) has a lower total time-cost and rental cost for the tested simulation task. It can be seen that using the cloud-based high-performance simulator, the user may accomplish massive simulation tasks with tiny rental costs of computing resources. This feature not only benefits individual researchers but also helps research facilities to carry out comprehensive studies on large-scale AC/DC hybrid power systems.

TABLE 1 RENTAL COST COMPARISON

| Device Type | Intel E5-2682 | NVIDIA P100 |
|---|---|---|
| Concurrent VSEs | 32 | 1 |
| Price | $115 / Week | $145 / Week |
| Time-costs (50μs integration step) | 0.53 h | 0.29 h |
| Rental Cost | $0.363 | $0.250 |

## V. CONCLUSION

In this paper, the design of high-performance power system simulator, CloudPSS, is introduced. Benefit from the proposed open cloud service integration framework, the automatic code generator, and the heterogeneous parallel computing technique, CloudPSS can provide highly efficient, flexible-to-use and cost-effective solutions for simulating large-scale power system and multi-scenario studies. It is worth noting that the proposed framework and the methodology of the automatic code generator are not limited to EMT simulations. They can be further extended to integrate power flow calculation, transient stability simulation and other applications in power system analysis in future research.

## REFERENCES

[1] J. Mahseredjian, V. Dinavahi, and J. A. Martinez, "Simulation Tools for Electromagnetic Transients in Power Systems: Overview and Challenges," *IEEE Trans. Power Deliv.*, vol. 24, no. 3, pp. 1657–1669, Jul. 2009.

[2] N. Watson and J. Arrillaga, *Power Systems Electromagnetic Transients Simulation*. IET, 2003.

[3] J. R. Marti, L. R. Linares, J. Calvino, H. W. Dommel, and J. Lin, "OVNI: an object approach to real-time power system simulators," in *1998 International Conference on Power System Technology, 1998. Proceedings. POWERCON '98*, 1998, vol. 2, pp. 977–981 vol.2.

[4] Z. Zhou and V. Dinavahi, "Parallel Massive-Thread Electromagnetic Transient Simulation on GPU," *IEEE Trans. Power Deliv.*, vol. 29, no. 3, pp. 1045–1053, Jun. 2014.

[5] J. K. Debnath, A. M. Gole, and W. k Fung, "Graphics Processing Unit based acceleration of Electromagnetic Transients Simulation," *IEEE Trans. Power Deliv.*, vol. PP, no. 99, pp. 1–1, 2015.

[6] Y. Song, Y. Chen, S. Huang, Y. Xu, Z. Yu, and J. R. Marti, "Fully GPU-based electromagnetic transient simulation considering large-scale control systems for system-level studies," *IET Transm. Distrib. Gener.*, vol. 11, no. 11, pp. 2840–2851, 2017.

[7] R. Singh *et al.*, "Using local grid and multi-core computing in electromagnetic transients simulation," *IPST 2013*, 2013.

[8] Y. Chen and V. Dinavahi, "Multi-FPGA digital hardware design for detailed large-scale real-time electromagnetic transient simulation of power systems," *IET Gener. Transm. Distrib.*, vol. 7, no. 5, pp. 451–463, May 2013.

[9] M. Armbrust *et al.*, "A view of cloud computing," *Commun. ACM*, vol. 53, no. 4, pp. 50–58, 2010.

[10] J. Turnbull, *The Docker Book: Containerization Is the New Virtualization*. James Turnbull, 2014.

[11] CloudPSS Group, "CloudPSS." [Online]. Available: http://www.cloudpss.net/.

[12] S. Crago *et al.*, "Heterogeneous cloud computing," in *Cluster Computing (CLUSTER), 2011 IEEE International Conference on*, 2011, pp. 378–385.

[13] J. A. Hollman and J. R. Marti, "Real time network simulation with PC-cluster," *IEEE Trans. Power Syst.*, vol. 18, no. 2, pp. 563–569, May 2003.

[14] NVIDIA, "CUDA Toolkit Documentation v9.0." NVIDIA, Sep-2017.

[15] J. C. Browne, "Framework for formulation and analysis of parallel computation structures," *Parallel Comput.*, vol. 3, no. 1, pp. 1–9, 1986.

[16] Y. Song, Y. Chen, Z. Yu, S. Huang, and L. Chen, "A fine-grained parallel EMTP algorithm compatible to graphic processing units," in *2014 IEEE PES General Meeting | Conference Exposition*, 2014, pp. 1–6.

[17] Y. Song, Y. Chen, S. Huang, Y. Xu, Z. Yu, and W. Xue, "Efficient GPU-Based Electromagnetic Transient Simulation for Power Systems With Thread-Oriented Transformation and Automatic Code Generation," *IEEE Access*, vol. 6, pp. 25724–25736, 2018.

[18] Z. Liu, Y. Song, Y. Chen, S. Huang, and M. Wang, "Batched Fast Decoupled Load Flow for Large-Scale Power System on GPU," in *2018 International Conference on Power System Technology (POWERCON 2018)*, Guangzhou, China, 2018, pp. 1–6.

[19] J. Cheng, M. Grossman, and T. McKercher, *Professional CUDA C Programming*. John Wiley & Sons, 2014.

[20] Alibaba Group, "Alibaba Cloud." [Online]. Available: https://www.alibabacloud.com.